\documentclass[twocolumn,aps,floatfix,superscriptaddress]{revtex4}
\pdfoutput=1 %This is for the ArXiv submission only
\usepackage{amsmath,amssymb,eucal,graphicx,float,epstopdf}

\usepackage[colorlinks=true, urlcolor=blue, anchorcolor=blue, citecolor=blue,filecolor=blue,linkcolor=blue,menucolor=blue]{hyperref}

\begin{document}
\title{Stochastic Dynamics of Growing Young Diagrams and Their Limit Shapes}

\author{P.~L.~Krapivsky}
\affiliation{Department of Physics, Boston University, Boston, MA 02215, USA}

\begin{abstract}
We investigate a class of Young diagrams growing via the addition of unit cells and satisfying the constraint that the height difference between adjacent columns $\geq r$. In the long time limit, appropriately re-scaled Young diagrams approach a limit shape that we compute for each integer $r\geq 0$.  We also determine limit shapes of `diffusively' growing Young diagrams satisfying the same constraint and evolving through the addition and removal of cells that proceed with equal rates. 
\end{abstract}

\maketitle

\section{Introduction}

Partitions of integers frequently appear mathematics, especially in combinatorics, number theory and group representations  \cite{Euler:book,HR18,R37,Apostol,Andrews,Tableaux,Mac99,Romik15}. Partitions are also increasingly popular in physics  \cite{Bethe36,Bohr,AK46,T49,Wu96,Weiss,Bhaduri04,Poland05,CMO,ND15,AO16}. By definition, a partition of a natural number $n$ is its representation as a sum 
\begin{equation}
\label{n:partition}
n=m_1+\ldots +m_k, \quad m_1\geq \cdots\geq m_k>0
\end{equation}
The total number of partitions of $n$ is denoted by $p(n)$. For instance,  $4=4, ~4=3+1, ~4=2+2, ~4=2+1+1$ and $4=1+1+1+1$ are all possible partitions of 4. Therefore $p(4)=5$. 

The study of partitions goes back to Leonhard Euler \cite{Euler:book}. One his result is the beautiful expression of the generating function encoding the sequence $p(n)$ through a neat infinite product 
\begin{equation}
\label{Euler}
\sum_{n\geq 0}p(n)\,q^n = \prod_{k\geq 1}\frac{1}{1-q^k}
\end{equation}
(It is convenient to set $p(0)=1$.) Using \eqref{Euler} one easily deduces the asymptotic behavior: $\ln p(n)\simeq 2\pi\sqrt{n/6}$ for $n\gg 1$. Hardy and Ramanujan \cite{HR18} derived a more precise asymptotic formula
\begin{equation}
\label{HR}
p(n)\simeq \frac{1}{4\sqrt{3}\,n}\,\exp\!\left[\pi\sqrt{\frac{2n}{3}}\right]
\end{equation}
Rademacher improved \eqref{HR} and derived an {\em exact} formula \cite{R37} for $p(n)$. His proof relies on the so-called circle method of Hardy, Littlewood, and Ramanujan together with marvelous properties of the Dedekind eta function which is ultimately related to the Euler's generating function \eqref{Euler}; see \cite{Apostol} for a pedagogical derivation of the exact Hardy-Ramanujan-Rademacher formula. 

\begin{figure}
\hspace*{-0.6cm}
\includegraphics[scale=0.33]{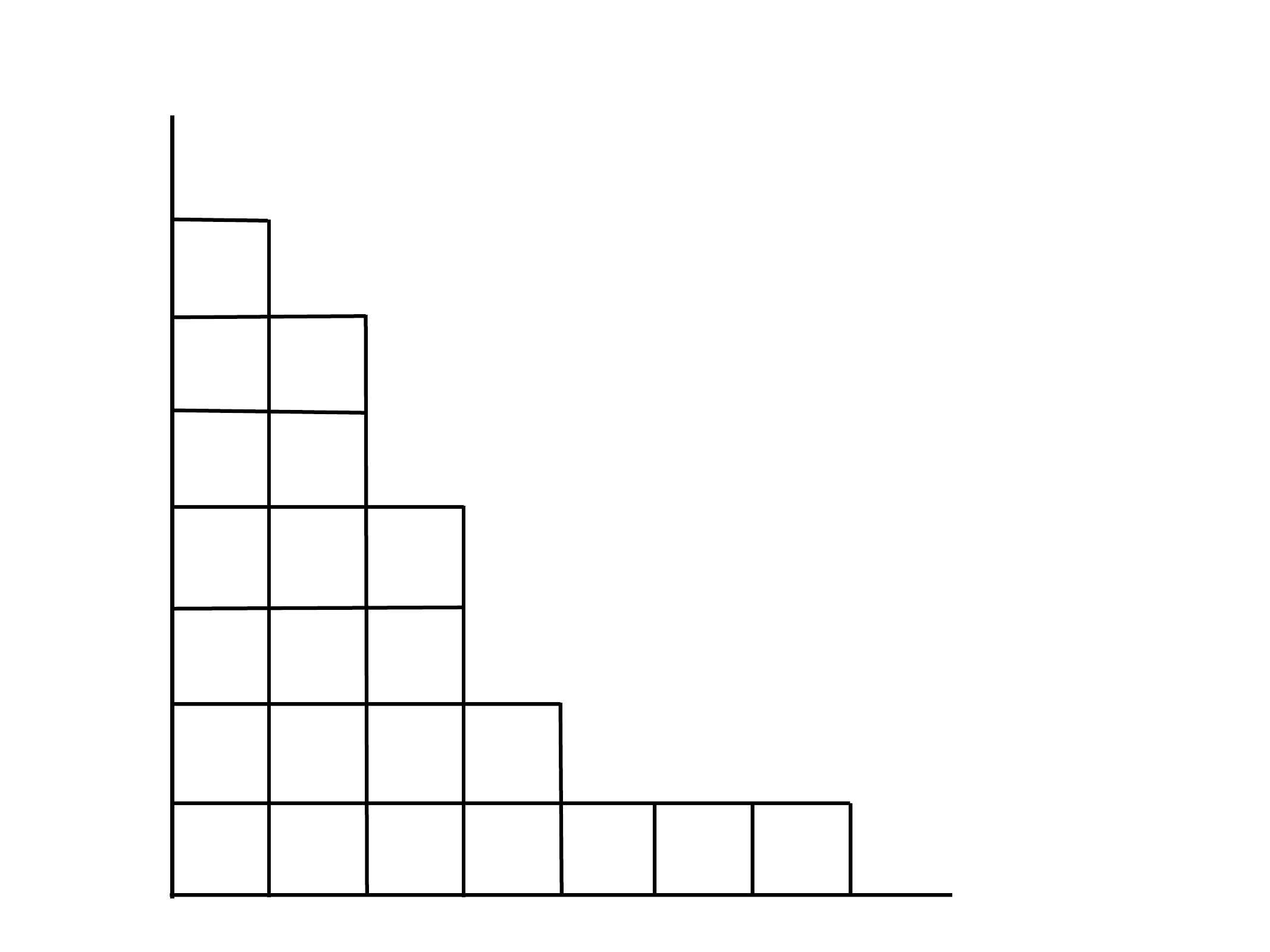}
\caption{A Young diagram of a partition of a positive integer $n$ is a diagram with $n$ boxes arranged in columns with non-increasing height. Shown is the Young diagram of the partition $22=7+6+4+2+1+1+1$.} 
\label{Fig:Young_2}
\end{figure}

One can think about partitions geometrically representing them by Young diagrams. This is illustrated in Fig.~\ref{Fig:Young_2}. The total number $\mathbb{Y}_2(n)$ of Young diagrams composed of $n$ elemental squares is $\mathbb{Y}_2(n)=p(n)$. Rather than fixing an area, one can impose other restrictions, e.g., one can consider Young diagrams that fit into an $a\times b$ box. The total number of such diagrams is 
\begin{equation}
\label{ab-box}
\mathbb{Y}(a,b) = \prod_{i=1}^a \prod_{j=1}^b  \frac{i+j}{i+j-1}
\end{equation}

The Young diagram is a two-dimensional (lattice) object, and it admits an obvious generalization to higher dimensions. The analog of Eq.~\eqref{Euler} is known in three (but not higher) dimensions \cite{Andrews,Mac99,Mac16}: 
\begin{equation}
\label{Mac:inf}
\sum_{n\geq 0}\mathbb{Y}_3(n)\,q^n = \prod_{k\geq 1}\frac{1}{(1-q^k)^k}
\end{equation}
where $\mathbb{Y}_3(n)$ is the total number of three-dimensional Young diagrams of `volume' $n$. This formula was discovered by MacMahon \cite{Mac16} who also found a beautiful formula for the total number $\mathbb{Y}(a,b,c)$ of Young diagrams that fit into an $a\times b\times c$ box generalizing \eqref{ab-box}:
\begin{equation}
\label{abc-box}
\mathbb{Y}(a,b,c) = \prod_{i=1}^a \prod_{j=1}^b \prod_{k=1}^c \frac{i+j+k-1}{i+j+k-2}
\end{equation}

Higher-dimensional partitions also appear in physics, e.g., in the context of the infinite-state Potts model \cite{Wu97}. Higher-dimensional partitions are still terra incognita. For instance, Eqs.~\eqref{Euler} and \eqref{Mac:inf}, as well as \eqref{ab-box} and \eqref{abc-box}, admit natural extensions to higher dimensions, but those formulas are erroneous, and the answers are unknown already in four dimensions.

The total number $p(n)$ of partitions rapidly grows with $n$, yet the Young diagrams look more and more alike in the $n\to\infty$ limit. To make this assertion precise, one must define the probability measure. The simplest choice is the {\em uniform} probability measure postulating that all $p(n)$ partitions of $n$ are equiprobable. The limit shape emerges after rescaling the coordinates  
\begin{equation}
\label{xy:def}
X = \frac{j}{\sqrt{n}}, \quad Y= \frac{m_j}{\sqrt{n}}
\end{equation}
and taking the $n\to\infty$ limit while keeping $X$ and $Y$ finite. The limit shape is given by  \cite{Temp} 
\begin{equation}
\label{uniform}
e^{-\lambda X} + e^{-\lambda Y} = 1, \quad \lambda= \frac{\pi}{\sqrt{6}}
\end{equation}
The amplitude $\lambda$ in \eqref{uniform} is fixed by the requirement that the area under the curve \eqref{uniform} is equal to $1$ thereby assuring that the area in the original coordinates is equal to $n$. There are various derivations \cite{Temp,VK,Martin,Vershik,VY} of the limit shape \eqref{uniform}. Most derivations rely on the Euler formula \eqref{Euler}, a few derivations use a variational approach \cite{SS,we}. Partitions with different probability measures have been also studied, e.g., the limit shape was determined \cite{VerKer} in the case of the Plancherel measure which naturally arises in the representation theory.

For three-dimensional Young diagrams of fixed large volume, the limit shape is known \cite{CK,OR} in the situation when the diagrams are taken with a uniform probability measure. The derivation in \cite{OR} uses the MacMahon formula \eqref{Mac:inf}. Three-dimensional Young diagrams equipped with the uniform probability measure and satisfying various constraints different from fixing the volume were studied in Refs.~\cite{CLP,CKP,OR07,KO,dFR}. For instance, the limit shape of three-dimensional Young diagrams fitting into large boxes was established in \cite{CLP}; the derivation relied, among other things, on the MacMahon formula \eqref{abc-box}.

Growing Young diagrams have been also investigated, see \cite{Rost,Liggett,KD90,barma,Ising_NNN,book,Ising_Area}. One postulates that new elemental squares are deposited stochastically in such a way that the growing object is always a proper Young diagram. More general stochastic rules allow both deposition and evaporation \cite{KD90,barma,we,Ising_NNN,book,Ising_Area}; again, the evolving object must remain a Young diagram. 
When only deposition events are allowed and occur at the same rate, the resulting process is known as the corner growth process in two dimensions. This process has numerous interpretations, e.g., one can think about it as the ``melting" of the Ising crystal at zero temperature \cite{KD90,barma,we,Ising_NNN,book,Ising_Area}. More precisely, one takes the Ising model with ferromagnetic nearest-neighbor interactions on the square lattice at zero temperature. The minority phase which initially occupies the positive quadrant constitutes the melting Ising crystal. If there is a (small) magnetic field favoring the majority phase, the zero-temperature spin-flip dynamics is equivalent to the corner growth process---only deposition events are possible. The growth is ballistic, $x\sim t$ and $y\sim t$. The limit shape is a parabola: $\sqrt{x}+\sqrt{y}=\sqrt{t}$. 

When the magnetic field is equal to zero, both deposition and evaporation events are possible and occur with equal rates. On average, the crystal exhibits a diffusive growth: $x\sim \sqrt{t}$ and $y\sim \sqrt{t}$. The limit shape and fluctuations are rather well-understood \cite{Ising_Area}. In three and higher dimensions, the analogous hyper-octant growth process has been studied \cite{Jason:12,Jason:13}, but the limit shapes are still unknown.

The analysis of evolving two-dimensional Young diagrams is simplified by mapping onto a one-dimensional totally asymmetric simple exclusion process. In the next section \ref{sec:GP}, we describe the mapping and outline how to use it to determine the limit shape for the corner growth process. The results in this section are known, but it is convenient to understand the mapping to the lattice gas in the simplest situation. Indeed, we use the same mapping in Secs.~\ref{sec:Fermi}--\ref{sec:general} to determine infinitely many limit shapes parametrized by a non-negative integer $r$, the minimal height difference between neighboring columns. The emergent lattice gases have hopping rules dependent on $r$.  In Sec.~\ref{sec:Fermi}, we consider the case of $r=1$ when growing partitions satisfy the constraint that all heights are different: $m_1> \cdots> m_k>0$. In Sec.~\ref{sec:general}, we investigate the general case when $m_j-m_{j+1}\geq r$. When deposition and evaporation occur with equal rates, Young diagrams grow on average in a diffusive manner. In Sec.~\ref{sec:DG}, we determine the corresponding limit shapes. The results are not fully explicit as for each $r\geq 1$  one must solve a second order nonlinear ordinary differential equation with one boundary found in the process of solution. The model with $r=1$ is more tractable as we show in Appendix \ref{ap:r=1}. In Appendix \ref{ap:Fluct}, we discuss fluctuations of the simplest quantities characterizing growing Young diagrams. For instance, the width of the Young diagram exhibits Gaussian fluctuations in the classical case of $r=0$; when $r=1$, the fluctuations of the width are non-trivial.

\section{Growing Young Diagrams and Lattice Gases}
\label{sec:GP}

\begin{figure}
\centering
\includegraphics[scale=0.44]{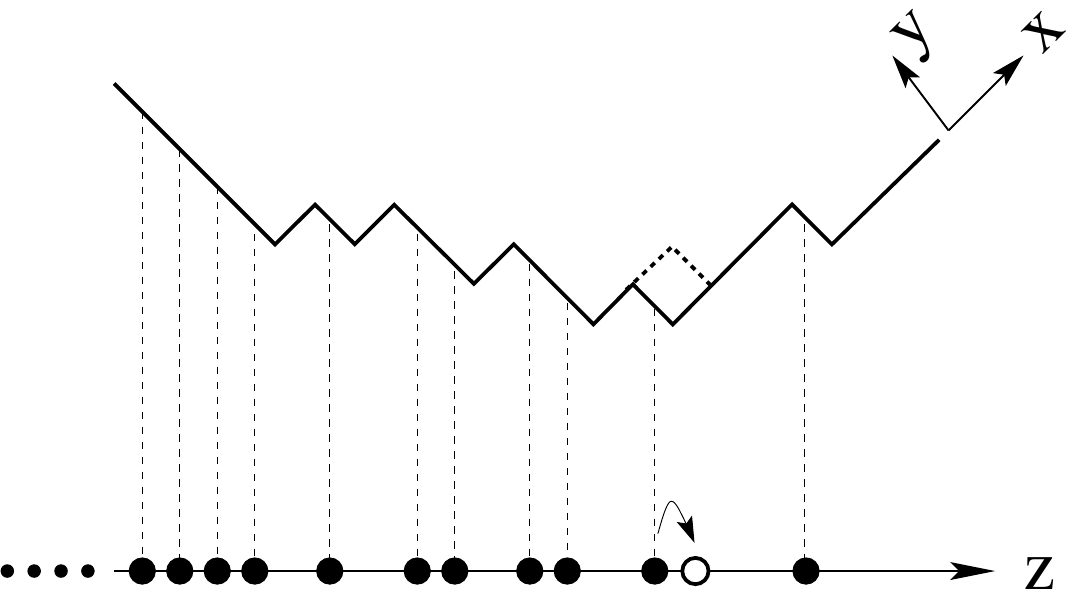}
\caption{The Young diagram from Fig.~\ref{Fig:Young_2} rotated by $\pi/4$. A deposition 
event is shown together with the corresponding hop of the particle in the underlying lattice gas.} 
\label{Fig:mapping}
\end{figure}

We consider growing two-dimensional Young diagrams and allow only deposition events if not stated otherwise. The deposition rules are different. In all cases, the analysis is simplified by mapping the growth process onto a one-dimensional lattice gas. The mapping is performed in two steps. First, we take the quadrant with the Young diagram at the corner and rotate counterclockwise by $\pi/4$ around the origin. Second, we project each $\diagdown$ segment to an occupied site ($\bullet$) and each $\diagup$ segment to an empty site ($\circ$) on the horizontal axis. This is illustrated in Fig.~\ref{Fig:mapping}. 

For instance, in the lattice gas representation the Young diagram from Fig.~\ref{Fig:Young_2} turns into 
\begin{equation*}
\ldots \bullet\bullet\bullet\circ\bullet\circ\bullet\bullet\circ\bullet\bullet\circ\bullet\circ\circ\circ\bullet\circ\circ\circ\ldots
\end{equation*}
The half-infinite fully occupied (unoccupied) part on the left (right) correspond to the part of the vertical (horizontal) axis in Fig.~\ref{Fig:Young_2}  that runs all the way to infinity. 

The simplest deposition procedure posits that whenever the deposition event is possible, it occurs at the same rate (set to unity without loss of generality). Emerging partitions depend on the realization of the stochastic deposition process (even the area is a stochastic variable). In the long time limit, however, the limit shape is reached after rescaling. We now explain how to determine the limit shape relying on the mapping onto a lattice gas \cite{Rost,Liggett,KD90,barma,Ising_NNN,Ising_Area,book,Ising_droplet}. In the lattice gas representation, the initial configuration --- an empty partition --- is a domain wall in which all the sites on the left (right) of the origin are occupied (unoccupied):
\begin{equation}
\label{IC}
\ldots \bullet\bullet\bullet\bullet\bullet\circ\circ\circ\circ\circ\ldots
\end{equation}
The only possible deposition event is at the corner. The corresponding partition $1=1$ is obtained from \eqref{IC} via the move 
\begin{equation}
\label{partition:1}
\ldots \bullet\bullet\bullet\bullet\bullet\circ\circ\circ\circ\circ\ldots \Longrightarrow
\ldots \bullet\bullet\bullet\bullet\circ\bullet\circ\circ\circ\circ\ldots
\end{equation}
in the lattice gas framework. Two possible deposition events can occur giving $2=2$ and $2=1+1$. In the lattice gas framework
\begin{equation}
\label{partition:2}
\bullet\bullet\bullet\bullet\circ\bullet\circ\circ\circ\circ  \Longrightarrow
\begin{cases}
\bullet\bullet\bullet\circ\bullet\bullet\circ\circ\circ\circ\\
\bullet\bullet\bullet\bullet\circ\circ\bullet\circ\circ\circ
\end{cases}
\end{equation}
These two partitions occur with the same probability reflecting that hopping events proceed with the same rate. There is still no difference with the equilibrium (uniform) measure. 

Two partitions which are possible outcomes of the process \eqref{partition:2} evolve with overall rate 2 and lead to
\begin{equation}
\label{partition:31}
\bullet\bullet\bullet\circ\bullet\bullet\circ\circ\circ\circ  \Longrightarrow
\begin{cases}
\bullet\bullet\circ\bullet\bullet\bullet\circ\circ\circ\circ\\
\bullet\bullet\bullet\circ\bullet\circ\bullet\circ\circ\circ
\end{cases}
\end{equation}
and 
\begin{equation}
\label{partition:32}
\bullet\bullet\bullet\bullet\circ\circ\bullet\circ\circ\circ  \Longrightarrow
\begin{cases}
\bullet\bullet\bullet\circ\bullet\circ\bullet\circ\circ\circ\\
\bullet\bullet\bullet\bullet\circ\circ\circ\bullet\circ\circ
\end{cases}
\end{equation}
Note that the partition $3=2+1$ which has the lattice gas representation $\bullet\bullet\bullet\circ\bullet\circ\bullet\circ\circ\circ$ occurs with probability 1/2 while other partitions, $3=1+1+1$ and $3=3$, occur with probability $1/4$ each. Thus different partitions may come with different weights hinting, correctly, that the limit shape of growing partitions is different from the equilibrium limit shape \eqref{uniform}.

The underlying lattice gas is known as a totally asymmetric simple exclusion process (TASEP). This is an exclusion process since every site can host at  most one particle, this exclusion process is `simple' since only nearest-neighbor hopping is allowed, and particles can hop only to the right (hence totally asymmetric). The rules of the TASEP are illustrated in Fig.~\ref{Fig:mapping}.

We now outline the derivation of the limit shape in the simplest case of unresticted growing partitions as we shall employ the same approach for other classes of growing partitions. The idea is to use the above lattice gas representation and to rely on a hydrodynamic description. This description ignores fluctuations, but it suffices for the derivation of the limit shape. For driven lattice gases (i.e., lattice gases with asymmetric hopping), the hydrodynamic description of the evolution of the average density $\rho(z,t)$ is based on a continuity equation 
\begin{equation}
\label{continuity}
\frac{ \partial \rho}{\partial t} +\frac{\partial J}{\partial  z} =0
\end{equation}
The current-density dependence is  \cite{Liggett,BE07} 
\begin{equation}
\label{current:TASEP}
J(\rho) = \rho(1-\rho)
\end{equation}
for the TASEP. Equations \eqref{continuity}--\eqref{current:TASEP} subject to the initial condition \eqref{IC}, equivalently
\begin{equation}
\label{IC:step}
\rho(z, 0) = 
\begin{cases}
1   & z<0\\
0   & z>0
\end{cases}
\end{equation}
admit a scaling solution: 
\begin{equation}
\label{scal-ansatz}
\rho(z,t) = N(Z), \qquad Z=\frac{z}{t}
\end{equation}
Plugging this ansatz into \eqref{continuity}--\eqref{current:TASEP} and solving the resulting equation yields the scaled density profile 
\begin{equation}
\label{nzt}
N(Z) = \left\{
\begin{array}{cl} \displaystyle
1         &~ Z<-1\\  \displaystyle
\tfrac{1}{2}(1- Z)   &~ |Z|<1\\
0         &~ Z>1
\end{array}
\right.
\end{equation}
This is an example of a rarefaction wave. Rarefaction waves are among the simplest solutions of hyperbolic partial differential equations, they shed light on the basic features of driven lattice gases (see \cite{book,BE07}). 

The limit shape is determined from the density through relation 
\begin{equation}
\label{yx}
y(x,t)=\int_{x-y}^\infty dz\,\rho(z,t)
\end{equation}
An exact discrete relation follows from Fig.~\ref{Fig:mapping}, and in the continuum limit it leads to \eqref{yx}. 
Rescaling the coordinates
\begin{equation}
\label{XY}
X=\frac{x}{t}\,,\quad Y=\frac{y}{t}
\end{equation}
we re-write \eqref{yx} as 
\begin{equation}
\label{YX}
Y=\int_{\max(X-Y,-1)}^1 dZ\,N(Z)
\end{equation}
Combining \eqref{nzt} and  \eqref{YX} we obtain an implicit equation for the limit shape
\begin{equation*}
4Y = \left\{
\begin{array}{cl} \displaystyle
1-2(X-Y)+ (X-Y)^2        &~ |X-Y|<1\\  \displaystyle
0                                  &~ X-Y>1
\end{array}
\right.
\end{equation*}
This equation can be recast into a manifestly symmetric form \cite{Rost}
\begin{equation}
\label{rost}
\sqrt{X}+\sqrt{Y}=1
\end{equation}
in the region $0<X,~Y<1$. 

\section{Growing Young Diagrams with unequal parts}
\label{sec:Fermi}

Partitions with the requirement that all parts are unequal were already studied by Euler \cite{Euler:book} who expressed the generating function for such partitions through an infinite product
\begin{equation}
\label{Euler:Fermi}
\sum_{n\geq 0} p_1(n)\,q^n = \prod_{k\geq 1}(1+q^k)
\end{equation}
Here the convention $p_1(0)=1$ is used again;  the index in the partition function $p_1(n)$ reminds about the requirement $m_j-m_{j+1}\geq 1$.  

For instance $6=6, ~6=5+1, ~6=4+2, ~6=3+2+1$ are the only possible partitions of 6 with unequal parts, so $p_1(6)=4$; the total number of unrestricted partitions of 6 is $p(6)=11$. Using \eqref{Euler:Fermi} and analyzing the $q\to 1$ behavior one can extract the asymptotic behavior:  $\ln p_1(n)\simeq \pi\sqrt{n/3}$ as $n\to \infty$.  A more comprehensive analysis \cite{Andrews} gives the Ramanujan asymptotic formula
\begin{equation*}
p_1(n)\simeq \frac{1}{4\cdot 3^{1/4}\,n^{3/4}}\,\exp\!\left[\pi\sqrt{\frac{n}{3}}\right]
\end{equation*}

The limit shape of partitions with unequal parts chosen uniformly among all $p_1(n)$ partitions has been established in Ref.~\cite{Vershik} using the generating function \eqref{Euler:Fermi}. In the re-scaled coordinates \eqref{xy:def} this limit shape reads
\begin{equation}
\label{uniform:Fermi}
e^{\lambda X} - e^{-\lambda Y} = 1, \quad \lambda= \frac{\pi}{\sqrt{12}}
\end{equation}

The limit shape \eqref{uniform} is symmetric with respect to the reflection $X\leftrightarrow Y$, and its span is infinite along both axes. (From \eqref{uniform} one finds that the span grows logarithmically, $X_*=Y_*=\frac{\sqrt{6}}{2\pi}\ln(n)$, so it diverges in the $n\to\infty$ limit.) The reflection symmetry is broken for the limit shape \eqref{uniform:Fermi} and the horizontal span of the partition is finite:
\begin{equation}
X\leq X_*= \frac{\sqrt{12}\,\ln 2}{\pi}
\end{equation}
In the original coordinates 
\begin{equation*}
j\leq j_*= \frac{\ln 2}{\pi}\,\sqrt{12n}
\end{equation*}
for $n\gg 1$. The maximal horizontal span is $j_\text{max}\approx \sqrt{2n}$, it arises for the least tilted partition with strictly decreasing heights: $j_\text{max}, ~j_\text{max} -1, \ldots, 1$. Almost all partitions with unequal parts are substantially more narrow:
\begin{equation*}
\frac{j_*}{j_\text{max}}= \frac{\sqrt{6}\ln 2}{\pi}= 0.54044463946673\ldots
\end{equation*}

We now turn to growing partitions with unequal parts. The first deposition event is the same as before, viz. \eqref{partition:1} in the lattice gas framework. The second deposition events is also unique: 
\begin{equation}
\label{partition:2F}
\bullet\bullet\bullet\bullet\circ\bullet\circ\circ\circ\circ  \Longrightarrow
\bullet\bullet\bullet\circ\bullet\bullet\circ\circ\circ\circ
\end{equation}
The third deposition event is described by \eqref{partition:31}, both outcomes occur with the same probability. Analyzing \eqref{partition:2F}, \eqref{partition:31}, and following deposition events one finds that the underlying lattice gas is a facilitated totally asymmetric simple exclusion process (FTASEP). The crucial difference from the TASEP is facilitation, a particle can hop only when it is pushed from the left (that is, its neighboring left site is occupied). 

\begin{figure}
\centering
\includegraphics[width=8cm]{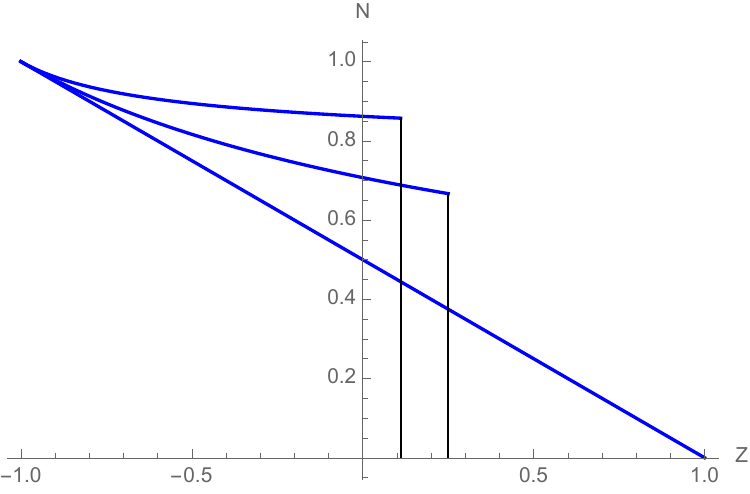}
\caption{The bottom curve is the classical rarefaction wave given by Eq.~\eqref{nzt}. The middle curve is the rarefaction wave \eqref{nzt:FASEP}, the top curve is the rarefaction wave given by Eq.~\eqref{nzt:r} with $r=4$. The last two rarefaction waves, and generally the rarefaction waves \eqref{nzt:r} with $r>0$, have a discontinuity on the right edge $Z=V(r)$; in the presented examples $V(1)=\frac{1}{4}$ and $V(4)=\frac{1}{9}$.}
\label{Fig:rarefaction}
\end{figure}

For the FTASEP we also use the continuity equation \eqref{continuity} on the hydrodynamic level. The FTASEP and closely related models were studied in the past \cite{SW98,KS99,AS00,LC03,SZL03,BM09,Alan10}, and  the dependence of the current from the density has been established
\begin{equation}
\label{current:FASEP}
J(\rho) = \frac{(1-\rho)(2\rho-1)}{\rho}
\end{equation}
To solve the continuity equation \eqref{continuity} with current \eqref{current:FASEP} and the initial condition \eqref{IC:step} we employ again the scaling ansatz \eqref{scal-ansatz}. One gets a rarefaction wave that has been found in Ref.~\cite{Alan10}:
\begin{equation}
\label{nzt:FASEP}
N(Z) = \left\{
\begin{array}{cl} \displaystyle
1         &~ Z<-1\\  \displaystyle
(2+ Z)^{-1/2}   &~ -1< Z <1/4\\
0         &~ Z > 1/4
\end{array}
\right.
\end{equation}
In contrast to shock waves, rarefaction waves usually exhibit a continuous (although not smooth) dependence on coordinate. The rarefaction wave \eqref{nzt:FASEP} is exceptional, the density jumps from $N=\frac{2}{3}$ at $Z=\frac{1}{4}-0$ to $N=0$ at $Z=\frac{1}{4}+0$ (see Fig.~\ref{Fig:rarefaction}). 

The limit shape is found by using \eqref{yx} which in the present case becomes
\begin{equation}
\label{YXZ:1}
Y=\int_{\max(X-Y,-1)}^{1/4} dZ\,N(Z)
\end{equation}
in the re-scaled coordinates \eqref{XY}. Combining \eqref{nzt:FASEP}  and \eqref{YXZ:1} we determine the limit shape 
\begin{equation}
\label{rost:1}
Y = 1 - 2\sqrt{X}
\end{equation}
Equation \eqref{rost:1} gives the non-trivial parabolic part of the limit shape in the region $0<X<\frac{1}{4},~0<Y<1$.  

\section{General Case}
\label{sec:general}

Here we look at $r-$partitions satisfying the requirement $m_j-m_{j+1}\geq r$, where $r$ is a fixed non-negative integer. (The height in the right-most column satisfies $m_k\geq 1$.) Unrestricted partitions are recovered when $r=0$; partitions with unequal parts correspond to $r=1$. 

There is an intriguing connection between $r-$partitions and systems of identical non-interacting quantum particles. Take a partition of $n$ and denote by $\nu_i$ the number of columns of height $i$. We have $n=\sum_{i\geq 1} i\nu_i$. We now interpret $n$ as the total energy of the system with levels labelled by $i$ occupied by $\nu_i$ particles; levels are assumed to be equidistant, so the energy at level $i$ is set equal to $i$. Unrestricted partitions correspond to bosons since they have arbitrary $\nu_i\geq 0$. Partitions with unequal parts have $n_i=0$ or $n_i=1$, so they correspond to fermions. Non-interacting quantum particles obeying exclusion statistics \cite{Haldane,Wu} can be related to $r-$partitions. The precise correspondence arises when $0<r<1$, but the results for the models with non-negative integer $r$  can be analytically continued \cite{CMO} to $0<r<1$. This connection with non-interacting quantum particles obeying exclusion statistics is mostly a motivation, particularly in our case of the growing partitions (i.e., non-conserved energy and the number of particles). 

The equilibrium case was studied in Refs.~\cite{SNM1,SNM2}. The generalization of \eqref{uniform} and \eqref{uniform:Fermi} reads 
\begin{equation}
\label{uniform:r}
e^{\lambda rX} - e^{-\lambda Y} = e^{\lambda (r-1)X} 
\end{equation}
with parameter $\lambda=\lambda(r)$ found \cite{SNM1,SNM2} by setting the area under the curve \eqref{uniform:r} to unity:
\begin{equation}
\lambda^2=\frac{\pi^2}{6}-\text{Li}_2(R)-\frac{r}{2}\,(\ln R)^2
\end{equation}
Here $R=R(r)$ is implicitly determined by $R+R^r=1$ and $\text{Li}_2(R)=\sum_{k\geq 1}k^{-2}R^k$ is the dilogarithm function.  For $r=0$ and $r=1$ one recovers the values given in \eqref{uniform} and \eqref{uniform:Fermi}; the next value is $\lambda(2)=\pi/\sqrt{15}$; etc.  

Let us look at growing $r-$partitions. The first unexplored case is $r=2$. In this model the first deposition event is described by \eqref{partition:1}, the second by \eqref{partition:2F}, the third deposition event is still unique
\begin{equation}
\label{partition:3F}
\bullet\bullet\bullet\circ\bullet\bullet\circ\circ\circ\circ  \Longrightarrow
\bullet\bullet\circ\bullet\bullet\bullet\circ\circ\circ\circ
\end{equation}
and only then there are two possible outcomes
\begin{equation}
\label{partition:4F}
\bullet\bullet\circ\bullet\bullet\bullet\circ\circ\circ\circ  \Longrightarrow
\begin{cases}
\bullet\circ\bullet\bullet\bullet\bullet\circ\circ\circ\circ\\
\bullet\bullet\circ\bullet\bullet\circ\bullet\circ\circ\circ
\end{cases}
\end{equation}
Analyzing \eqref{partition:1}, \eqref{partition:2F}, \eqref{partition:3F} and \eqref{partition:4F} we see that the underlying lattice gas is a facilitated asymmetric simple exclusion process FTASEP(2) where a particle can hop only when it is pushed from the left by two adjacent particles. 

Generally growing $r-$partitions are mapped onto the FTASEP($r$), an exclusion process in which the push by $r$ adjacent particles from the left is required for the hop to the neighboring empty site on the right. Remarkably, the current is known for all such processes \cite{find:RP}:
\begin{equation}
\label{current:r}
J(\rho) = \frac{(1-\rho)[(r+1)\rho-r]}{r\rho +1-r}
\end{equation}

\begin{figure}
\centering
\includegraphics[width=8cm]{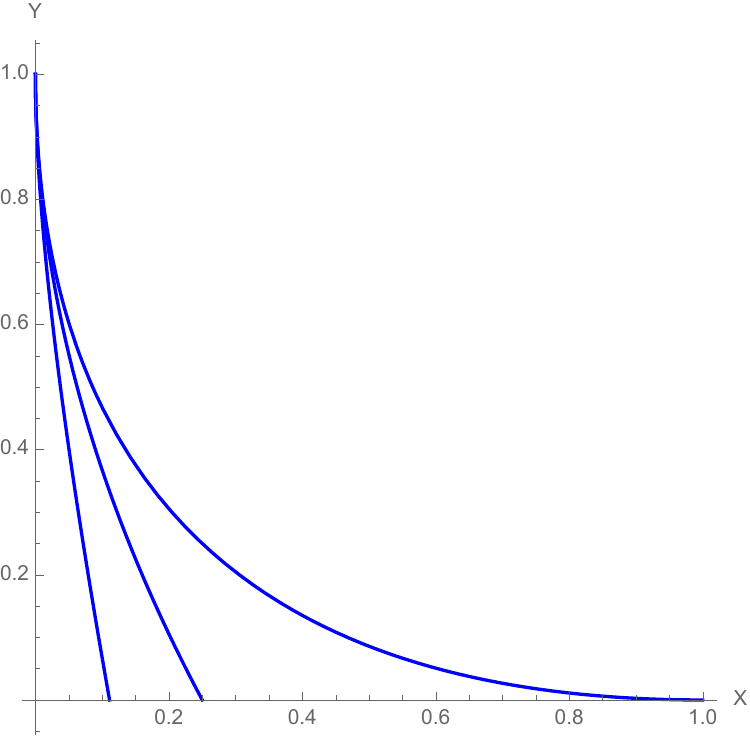}
\caption{Top to bottom: The limit shapes with $r=0, ~1, ~4$. The limit shape for arbitrary $r$ is given by Eq.~\eqref{rost:r}. In the simplest cases $r=0$ and $r=1$, the limit shapes also appear as \eqref{rost} and \eqref{rost:1}.}
\label{Fig:growth}
\end{figure}

Thus, we ought to solve the continuity equation \eqref{continuity} with current given by \eqref{current:r} subject to the initial condition \eqref{IC:step}. The non-trivial part of the corresponding rarefaction wave has again the scaling form \eqref{scal-ansatz}. Solving the ordinary differential equation for $N(Z)$ we determine the non-trivial part of the density
\begin{equation}
\label{nzt:r}
N(Z) = 1-r^{-1}+r^{-1}[1+r+ rZ]^{-1/2}
\end{equation}
which is valid in the region
\begin{equation}
\label{Vr}
-1 < Z < V(r), \qquad V(r) = \frac{1}{\big(\sqrt{r}+1\big)^2}
\end{equation}
Outside the region \eqref{Vr}, the density profile remains unperturbed: $N(Z)=1$ for $Z<-1$ and $N(Z)=0$ for $Z>V(r)$. The non-trivial part of the limit shape is surprisingly simple:
\begin{equation}
\label{rost:r}
Y = 1 - 2\sqrt{X}-(r-1)X, \quad 0<X<V(r)
\end{equation}
A few of these non-trivial parts of limit shapes are plotted in Fig.~\ref{Fig:growth}. Geometrically, each is a part of a parabola. The area under the parabola \eqref{rost:r} is
\begin{equation}
A(r) = \frac{3\sqrt{r}+1}{6\big(\sqrt{r}+1\big)^3}
\end{equation}
The area $A(r)$ is a decreasing function of $r$, see also Fig.~\ref{Fig:growth}.

In the original coordinates
\begin{equation}
\label{xy:max}
y_\text{max}=t, \qquad x_\text{max}= V(r)t = \frac{t}{\big(\sqrt{r}+1\big)^2}
\end{equation}
and the area is $A(r) t^2$ in the leading order.

\section{Diffusive Growth}
\label{sec:DG}

In the previous sections, we have studied strictly growing partitions (only deposition events were allowed). We have investigated different types of partitions: arbitrary partitions, partitions with unequal parts, and generally partitions with height difference $\geq r$. In all examples, the growth is ballistic, see \eqref{xy:max}. 

\begin{figure}
\vspace*{-0.2cm}
\hspace*{-0.44cm}
\includegraphics[scale=0.36]{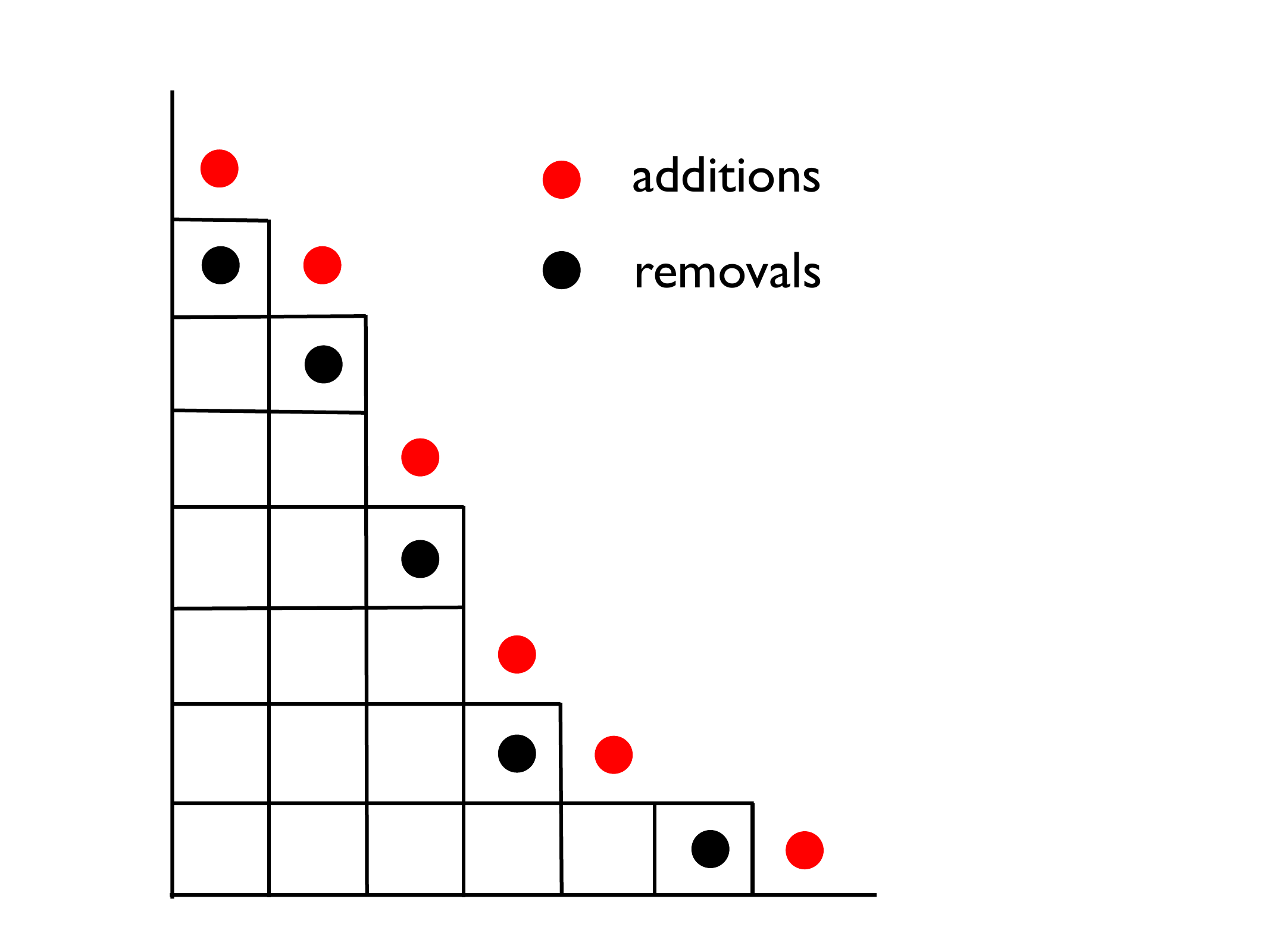}
\caption{An illustration of a dynamics acting on the set of arbitrary partitions in which additions and removals of squares occur with the same rates (set equal to unity). Shown is the partition $21=7+6+4+2+1+1$ together with six spots for the addition of squares and five spots for removal. There is always one more spot for addition than for removal, so the average area is $\langle S\rangle = t$ implying that the typical size grows diffusively as $\sqrt{t}$. } 
\label{Fig:Diff_growth}
\end{figure}

One can allow both additions and removals of squares, requiring that the evolving Young diagram remains the Young diagram of the prescribed type. If the addition rate exceeds the evaporation rate, the growth remains ballistic, and the limit shapes are the same as before up to a scaling factor. A qualitatively different diffusive growth occurs if additions and removals of squares proceed with equal rates. 

Figure \ref{Fig:Diff_growth} illustrates the process with equal rates of additions and removals in the case of arbitrary partitions. The number of positions where new squares can be added always exceeds by one the number of positions from which squares can be removed. Hence the average area increases linearly in time:
\begin{equation}
\label{Sav}
\langle S\rangle = t
\end{equation}

In the case of arbitrary partitions the above evolution process maps onto the symmetric simple exclusion process (SSEP) for which the diffusion equation, $\rho_t=\rho_{zz}$, provides the hydrodynamic description. Solving this equation subject to the initial condition \eqref{IC:step} one gets 
\begin{equation}
\label{Nz:SEP}
\rho(z,t)=\tfrac{1}{2}\text{Erfc}(\zeta), \quad \zeta = \frac{z}{\sqrt{4t}}
\end{equation}
which in conjunction with \eqref{yx} give the limit shape. 

\begin{figure}
\vspace*{-0.2cm}
\hspace*{-0.66cm}
\includegraphics[scale=0.36]{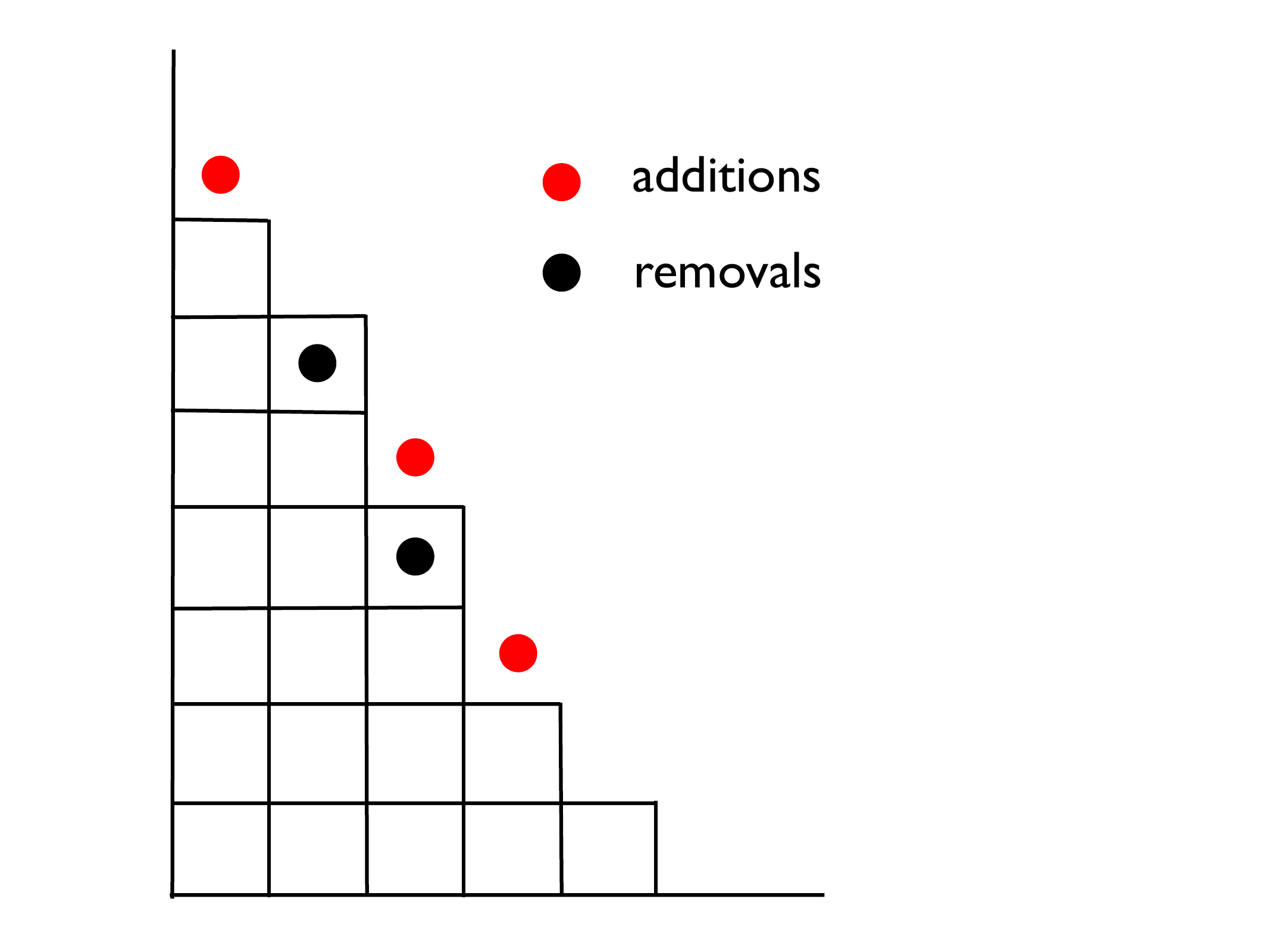}
\caption{An illustration of partitions with unequal parts evolving via additions and removals that proceed with equal rates. Shown is the partition of $20=7+6+4+2+1$ that has has three spots for addition and two spots for removal. In this example $m_5=1$ and $m_6=0$, so the right-most square cannot be removed since $m_5-m_6\geq 1$ is required.} 
\label{Fig:YD-Fermi}
\end{figure}

A diffusive growth of partitions with unequal parts (Fig.~\ref{Fig:YD-Fermi}) maps onto the facilitated symmetric simple exclusion process (FSSEP) in which the hopping is facilitated (caused by the nearest neighbor) and symmetric. The hydrodynamic description of the FSSEP is provided by a partial differential equation (PDE) 
\begin{equation}
\label{DE:FSEP}
\frac{ \partial \rho}{\partial t} =\frac{\partial}{\partial  z}\!\left(\frac{1}{\rho^2}\, \frac{\partial \rho}{\partial  z}\right)
\end{equation}
This diffusion equation is non-linear since the diffusion coefficient $D(\rho)=\rho^{-2}$ depends on the density. As usual, the hydrodynamic description is applicable when the characteristic spatial and temporal scales greatly exceed the microscopic scales, i.e., the lattice spacing and the inverse hopping rates which we have set to unity. Aside from this generic caveat, for the FSSEP the hydrodynamic description \eqref{DE:FSEP}  is applicable only when density is sufficiently large, $\frac{1}{2}\leq \rho\leq 1$; in the low-density regime, $0<\rho<\frac{1}{2}$, the FSSEP quickly reaches a jammed state and the evolution ceases. In a jammed state adjacent particles separated by at least one vacancy. 

The expression $D(\rho)=\rho^{-2}$ for the density-dependent diffusion coefficient has been extracted from the diffusion coefficient characterizing a repulsion process  \cite{RP13}. At first sight, these two processes are very different, e.g., the repulsion process has a well-defined hydrodynamic behavior in the entire range $0\leq \rho\leq 1$. In the $\frac{1}{2}\leq \rho\leq 1$ range, however, both the repulsion process and the FSSEP have an identical structure of the equilibrium states. Therefore we can use the already known \cite{RP13} expression of the diffusion coefficient for the repulsion process. The expression $D(\rho)=\rho^{-2}$ for the diffusion coefficient has been also recently derived \cite{Poincare20} fully in the realm of the FSSEP. 

The solution of Eq.~\eqref{DE:FSEP} subject to the initial condition \eqref{IC:step} has a self-similar form
\begin{equation}
\label{Nz:scaling}
\rho(z,t)=N(\zeta), \quad \zeta = z\sqrt{\frac{\pi}{4t}}
\end{equation}
By inserting the ansatz \eqref{Nz:scaling} into the governing PDE, we reduce Eq.~\eqref{DE:FSEP} to an ordinary differential equation
\begin{equation}
\label{Nz:ODE}
(N^{-2}N')' + \frac{2\zeta}{\pi}\, N' = 0
\end{equation}
where prime denotes differentiation with respect to $\zeta$. We must solve \eqref{Nz:ODE} in the region $-\infty<\zeta<v$. 

The boundary condition at $\zeta\to -\infty$ is 
\begin{equation}
\label{Nz:left}
N(-\infty) = 1
\end{equation}

The density at the right boundary is the minimal allowed density where the hydrodynamic description holds:
\begin{subequations}
\begin{equation}
\label{N:right}
N(v) = \frac{1}{2}
\end{equation}
We need an additional boundary condition since the position of the right boundary, $\zeta=v$, is unknown. The current through it is 
$-D(\rho)\rho_z= -N^{-2}N' \sqrt{\frac{\pi}{4t}}$. On the other hand, the current is $N\frac{d}{dt}v\sqrt{4t/\pi}=Nv/\sqrt{\pi t}$. Equating these two expressions and using \eqref{N:right} we get
\begin{equation}
\label{N-prime:right}
N'(v) = -\frac{v}{4\pi}
\end{equation}
\end{subequations}

The non-linear ordinary differential equation \eqref{Nz:ODE} with unknown boundary can be solved analytically. The trick is to map the FSSEP into the SSEP. More precisely, the FSSEP with domain wall initial condition \eqref{IC:step} can be mapped into the half-SSEP with a localized source at the boundary as we show in Appendix \ref{ap:r=1}. This latter problem is analytically tractable. To determine the density $N(\zeta)$, one must transform both the spatial variable and the density through the spatial variable and the density in the SSEP problem. Completing this program yields the solution in a parametric form 
\begin{subequations}
\begin{align}
\label{N-diff:sol}
&N(\zeta) = \frac{1}{1+\text{Erfc}(\xi)}\\
\label{parameters-diff:sol}
&\zeta = e^{-\xi^2}-\sqrt{\pi}\,\xi\left[1+\text{Erfc}(\xi)\right]
\end{align}
\end{subequations}
The derivation of \eqref{N-diff:sol}--\eqref{parameters-diff:sol} is given in Appendix \ref{ap:r=1}. 

\begin{figure}
\centering
\includegraphics[width=8cm]{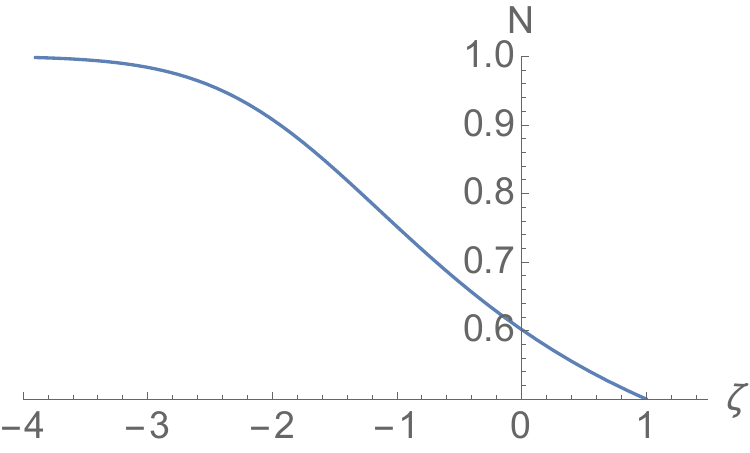}
\caption{The scaled density profile \eqref{N-diff:sol}--\eqref{parameters-diff:sol}.}
\label{Fig:N1-diff}
\end{figure}

The density profile \eqref{N-diff:sol}--\eqref{parameters-diff:sol} describes the density on the half-line $\zeta\leq v=1$ (see Fig.~\ref{Fig:N1-diff}); for $\zeta>1$, the density vanishes: $N=0$. (We have included $\pi$ into the definition of the scaling variable $\zeta$ in Eq.~\eqref{Nz:scaling} to ensure that $v=1$.) Therefore in the original coordinates, the average position of the right-most particle (equivalently, the average width of the diffusively growing partition with unequal parts) is
\begin{equation}
\label{width:1}
\langle w_t\rangle = \sqrt{\frac{4t}{\pi}}
\end{equation}

\begin{figure}
\centering
\includegraphics[width=8cm]{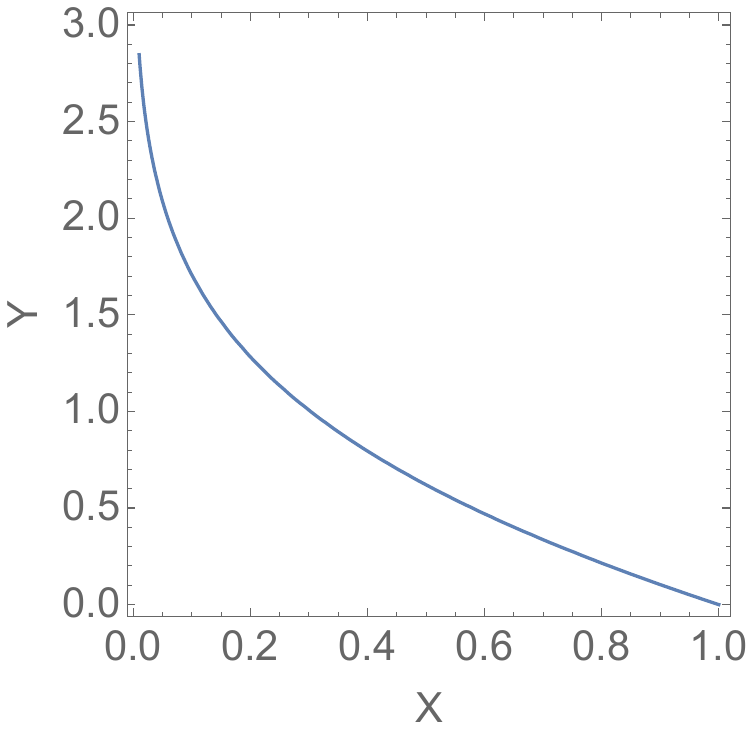}
\caption{The limit shape in the case of diffusively growing partitions with unequal parts ($r=1$). The limit shape is determined by Eq.~\eqref{LS:diff-XY}.}
\label{Fig:Diff-LS-new}
\end{figure}

In the re-scaled coordinates
\begin{equation}
\label{XY-diff}
X=x \sqrt{\frac{\pi}{4t}}\,,\quad Y=  y\sqrt{\frac{\pi}{4t}}
\end{equation}
the limit shape is determined via
\begin{equation}
\label{LS:diff-1}
Y=\int_{X-Y}^1 d\zeta\,N(\zeta)
\end{equation}
Using Eqs.~\eqref{N-diff:sol}--\eqref{parameters-diff:sol}, one computes the integral and recasts \eqref{LS:diff-1} into $Y=\sqrt{\pi}\,\xi(X-Y)$. This is further simplified, with help of Eq.~\eqref{parameters-diff:sol}, to a concise equation for the limit shape:
\begin{equation}
\label{LS:diff-XY}
e^{-Y^2/\pi}-Y\, \text{Erfc}\big(Y/\sqrt{\pi}\big)=X
\end{equation}
This limit shape is shown in Fig.~\ref{Fig:Diff-LS-new}.

There is one more spot for addition than for removal of squares, so the average area is again given by Eq.~\eqref{Sav}. This growth law also follows from the diffusion equation \eqref{DE:FSEP} thereby providing a consistency check. Indeed, the average area varies with unit rate:
\begin{equation}
\frac{d \langle S\rangle}{dt} = - \int_{-\infty}^{\sqrt{4t/\pi}}dz\,\rho^{-2}\rho_z=-\int_{1}^{\frac{1}{2}}\frac{dN}{N^2} = 1 
\end{equation}

Generally for the diffusive growth of partitions satisfying the requirement $m_j-m_{j+1}\geq r$ the governing PDE is again a non-linear diffusion equation
\begin{equation}
\label{DEr}
\frac{ \partial \rho}{\partial t} =\frac{\partial}{\partial  z}\!\left[\frac{1}{(r\rho -r+1)^2}\, \frac{\partial \rho}{\partial  z}\right]
\end{equation}
with density-dependent diffusion coefficient. 
The hydrodynamic description is applicable in the density range $\frac{r}{r+1}\leq \rho\leq 1$, and the diffusion coefficient 
is again established through the relation to the generalized repulsion process \cite{RP13,Ising_NNN}. The solution has a self-similar form \eqref{Nz:scaling}. The scaling function satisfies
\begin{equation}
\label{Nzr}
[(rN-r+1)^{-2}N']' + \frac{2\zeta}{\pi}\, N' = 0
\end{equation}
The boundary conditions on the right edge are
\begin{equation}
\label{Nr:right}
N(v) =  \frac{r}{r+1}\,, \quad N'(v) = -\frac{2v}{\pi (r+1)^3}
\end{equation}
Analytical expressions for $v=v(r)$ are unknown for all $r\geq 2$.

\section{Concluding Remarks}

We have computed limit shapes characterizing growing two-dimensional Young diagrams parametrized by a non-negative integer $r$, the minimal difference between the heights of adjacent columns. Infinitely many limit shapes were also computed \cite{Ising_NNN} for the melting Ising crystals on the square lattice with ferromagnetic spin-spin interactions; these limit shapes are parametrized by the range of interaction.  

It would be interesting to study the three-dimensional growing Young diagrams satisfying the constraint that for any $(i,j)$ with $m_{i,j}>0$, the heights of neighboring columns are smaller at least by $r$, i.e., $m_{i,j}-m_{i+1,j}\geq r$ and  $m_{i,j}-m_{i,j+1}\geq r$. Even for arbitrary Young diagrams,  the mapping of the growing interface in three dimensions onto a two-dimensional lattice gas has not led so far to a scheme allowing to extract a limit shape. One can also try to guess a PDE for the limit shape $z=z(x,y,t)$ satisfying proper symmetry conditions. This approach has led to a prediction \cite{Jason:12,Jason:13} tantalizingly close to simulation results. The PDE in the three-dimensional situation admits a generalization to an arbitrary dimension. It would be interesting to guess similar PDEs for restricted classes of Young diagrams.

\bigskip 

{\bf Acknowledgments}. I am grateful to K. Mallick for useful discussions. I also benefitted from the correspondence with G. Barraquand, I. Corwin, and T. Sasamoto. 

\appendix

\section{ Diffusively growing partitions with unequal parts}
\label{ap:r=1}

First, we describe the mapping \cite{BBC} of the FTASEP on the half-TASEP with a source at the boundary. The sites in the TASEP correspond to adjacent particles in the FTASEP. In the TASEP, the site is empty ($\square$) if adjacent particles are nearest neighbors;  if there is a vacancy between adjacent particles, the site is occupied ($\blacksquare$). Here is an illustration of the evolution (time goes from top to bottom)
\begin{equation}
\label{evolution}
\begin{split}
&\ldots \bullet\bullet\bullet\bullet\bullet\bullet\circ\circ\circ\ldots      
\qquad   \ldots\square\square\square\square\square \\
&\ldots\bullet\bullet\bullet\bullet\bullet\circ\bullet\circ\circ\ldots  
\qquad  \ldots\square\square\square\square\blacksquare    \\
&\ldots\bullet\bullet\bullet\bullet\circ\bullet\bullet\circ\circ\ldots   
\qquad  \ldots\square\square\square\blacksquare\square  \\
&\ldots\bullet\bullet\bullet\bullet\circ\bullet\circ\bullet\circ\ldots 
\qquad \ldots\square\square\square\blacksquare\blacksquare \\
&\ldots\bullet\bullet\bullet\circ\bullet\bullet\circ\bullet\circ\ldots 
\qquad \ldots\square\square\blacksquare\square\blacksquare \\
&\ldots\bullet\bullet\circ\bullet\bullet\bullet\circ\bullet\circ\ldots 
\qquad \ldots\square\blacksquare\square\square\blacksquare \\
&\ldots\bullet\bullet\circ\bullet\bullet\circ\bullet\bullet\circ\ldots 
\qquad  \ldots\square\blacksquare\square\blacksquare\square 
\end{split}
\end{equation}
The lattice model shown on the right is the half-TASEP with particles hopping to the left and new particles entering the right-most site (both processes occur with unit rate).  The mapping is one-to-one since starting with initial condition \eqref{IC} adjacent particles in the FTASEP remain separated by at most one vacancy. The FTASEP process begins with the domain wall initial condition \eqref{IC} that corresponds to the empty half-line in the half-TASEP, see the top line in \eqref{evolution}. 

The mapping of the FSSEP to the half-SSEP with a source at the boundary is identical. In both cases, a simple but crucial observation evident from the example \eqref{evolution} is that the displacement $w_t$ of the right-most particle for the original process (FTASEP or FSSEP)  is the same as the total number of particles in the process obtained by mapping (half-TASEP or half-SSEP). 

For the half-SSEP starting with an empty half-line and driven by a source at the boundary, one can solve the exact lattice equations for the density  \cite{Gunter,PK_input}, this is even possible for an arbitrary ratio of the input rate to the hopping rate. In long time limit, however, we can set the density at the boundary to unity and deduce the leading behavior from the simpler hydrodynamic framework
\begin{equation}
\label{nu:eq}
\nu_t=\nu_{\ell\ell}, \quad \nu(0,t>0)=1, \quad \nu(\ell>0,0)=0
\end{equation}
describing the evolution of the density $\nu(\ell,t)$ in the half-SSEP.  Solving \eqref{nu:eq} one finds
\begin{equation}
\label{N-ell:SEP}
\nu(\ell,t)=\text{Erfc}(\xi), \qquad \xi=\frac{\ell}{\sqrt{4t}}
\end{equation}
Using the identification of the displacement $w_t$ of the right-most particle in the FSSEP with the total number of particles in the half-SSEP, we get
\begin{equation*}
\langle w_t\rangle = \int_0^\infty d\ell\,  \nu(\ell,t)=
\sqrt{\frac{4t}{\pi}}
\end{equation*}
as was stated in \eqref{width:1}. 

The mapping between the FSSEP and the half-SSEP tells us that the distance $\ell$ from the boundary in the half-SSEP corresponds to the distance 
\begin{equation}
\label{L-diff}
L = \int_0^\ell du\left[1+\nu(u,t)\right]
\end{equation}
from the right-most particle in the FSSEP. Computing the integral in \eqref{L-diff} yields
\begin{equation}
\label{L-xi}
L\sqrt{\frac{\pi}{4t}}=1-e^{-\xi^2}+\xi[1+\text{Erfc}(\xi)]
\end{equation}
The spatial coordinate in the FSSEP is $z=\langle w_t\rangle -L$, so the re-scaled spatial coordinate is $\zeta=1-L\sqrt{\frac{\pi}{4t}}$. This together with \eqref{L-xi} give Eq.~\eqref{parameters-diff:sol} relating $\zeta$ and $\xi$. 

The mapping of the FSSEP into the half-SSEP implies the relation $\rho=1/(1+\nu)$ between the corresponding densities. This relation together with \eqref{N-ell:SEP} yield the parametric representation \eqref{N-diff:sol} of the density. 

\section{Fluctuations}
\label{ap:Fluct}

Understanding of fluctuations of growing interfaces, particularly one-dimensional interfaces, has greatly improved over the last 30 years \cite{HZ95,spohn,Ivan_rev,HT15}. Growing arbitrary partitions have played a crucial role as the first example where fluctuations have been understood \cite{Johan}. Using the mapping onto the TASEP one can explore fluctuations in the latter framework. For the TASEP starting with the initial condition \eqref{IC}, the quantity that has been particularly well explored is the total number of particles $P_t$ entering the initially empty half-line during time interval $(0,t)$. It was shown \cite{Johan} that 
\begin{equation}
\label{Nt}
P_t =\frac{t}{4}+t^{1/3} \mathcal{F}_{GUE}
\end{equation}
where $\mathcal{F}_{GUE}$ is the Tracy-Widom GUE distribution; the GUE abbreviation reflects that it arises in the Gaussian unitary ensemble of random matrices. (The interface intersects the diagonal at the point $(P_t, P_t)$, so the fluctuations of the random quantity $P_t$ are directly related to fluctuations of the interface in the $(1,1)$ direction.) 

In the case of growing partitions with unequal parts, and generally, for models with $r>0$, random quantities like the width of the partition already exhibit intriguing and usually unknown behaviors. First, we show that when $r=0$, fluctuations of the width and height are Gaussian. This is evident in the lattice gas representation. For instance, the width is the displacement of the right-most particle which is independent of other particles, it merely hops to the right with unit rate. Therefore the width $w_t$ has the Poisson distribution
\begin{equation}
\text{Prob}(w_t=m)=\frac{t^m}{m!}\,e^{-t}
\end{equation}
which is asymptotically Gaussian with fluctuation on the scale $t^{1/2}$. Similarly, the height is the displacement of the left-most vacancy, so it has the same Poisson distribution.

Consider now strictly growing partitions with unequal parts, $r=1$. In this case, fluctuations of the width $w_t$ have been explored via the mapping into the half-TASEP described in Appendix \ref{ap:r=1}. Since the random quantity $w_t$ is equal to the (growing) total number of particles in  the half-TASEP, one sees the analogy with the quantity $P_t$ for the TASEP, and hence one expects that fluctuations scale as $t^{1/3}$. This is true, but they follow \cite{BBC} the GSE Tracy-Widom distribution related to the Gaussian symplectic ensemble of random matrices:
\begin{equation}
\label{wt}
w_t =\frac{t}{4}+C_1\,t^{1/3} \mathcal{F}_{GSE}
\end{equation}

The same $\mathcal{F}_{GSE}$ appears in other growth processes in half-line \cite{BR-half,BR-rev,TS04}, while the $\mathcal{F}_{GUE}$ distribution describes fluctuations of the leading particle in a process studied in Ref.~\cite{BC}. The behavior of the random quantity $w_t$ for models with $r\geq 2$ is unknown. If the qualitative behavior as in the $r=1$ case and only the magnitude of fluctuations is affected, 
\begin{equation}
\label{wt-r}
w_t = \frac{t}{\big(\sqrt{r}+1\big)^2} +C_r\,t^{1/3} \mathcal{F}_{GSE}
\end{equation}

Finally, let us discuss fluctuations for diffusively growing partitions. In the case of unrestricted partitions, the mapping into the SSEP provides a significant simplification.  Fluctuations of the width and height are easy to understand. The average displacement of the right-most particle can be estimated from the criterion
\begin{equation}
\label{wt:crit}
\int_{\langle w_t\rangle}^\infty dz\,\rho(z,t) \sim 1
\end{equation}
Combining \eqref{Nz:SEP} and \eqref{wt:crit} one gets $\langle w_t\rangle \simeq \sqrt{2t \ln t}$. One can heuristically estimate the variance of the width, $\langle w_t^2\rangle-\langle w_t\rangle^2$, by arguing that it scales as the square of the average gap $\langle g_t\rangle$ between the right-most particle and the preceding particle. This gap can be estimated from the criterion $\int_{\langle w_t\rangle - \langle g_t\rangle}^{\langle w_t\rangle} dz\,\rho(z,t) \sim 1$ to give
\begin{equation}
\label{var:arbitrary}
\langle w_t^2\rangle-\langle w_t\rangle^2 \sim \frac{t}{\ln t}
\end{equation}
More precise results are available in the situation when particles undergo Brownian motions \cite{SS07}, while the relevant case of the SSEP is studied in \cite{KT16}. 

In the case of partitions with unequal parts, we use again the mapping into the half-SSEP with a source. In addition to the average position of the width, Eq.~\eqref{width:1}, the variance has been determined \cite{Gunter,PK_input}. It also exhibits a diffusive growth. The ratio of the variance to the average, the Fano factor, is asymptotically
\begin{equation}
\label{var:strict}
\frac{\langle w_t^2\rangle-\langle w_t\rangle^2}{\langle w_t\rangle}= 3-\sqrt{8}
\end{equation}
Thus $w_t =  2\sqrt{t/\pi} + t^{1/4}\mathcal{W}$ with a certain (apparently unknown) random distribution $\mathcal{W}$. 

The area is a basic characteristic of the Young diagram supplementing height and width. For unrestricted diffusively growing partitions, fluctuations of the area have been probed in \cite{Ising_Area}. These fluctuations are strongly non-Gaussian as manifested by the growth of the cumulants: $\langle S^p\rangle_c = A_p t^{(p+1)/2}$. For $p\leq 4$, the amplitudes $A_p$ have been determined analytically \cite{Ising_Area} using the perturbative approach \cite{KM_var}. For diffusively growing partitions with unequal parts, the computation of the cumulants of the area beyond $\langle S\rangle=t$ seems challenging.  The perturbative approach \cite{KM_var} is efficient only for the lattice gases with a constant diffusion coefficient. The mapping on the half-SSEP may help if one would find a simple description of the area in the realm of the half-SSEP.


\begin{thebibliography}{99}
\bibitem{find:RP}
        Equation \eqref{current:r} is difficult to extract from  \cite{SW98,KS99,AS00,LC03,SZL03,BM09,Alan10}. 
        The same current appears in the proper density range for repulsion processes: The top line of 
        Eq. (21) from \cite{RP13}, with $m\to r$ and $\rho\to 1-\rho$,  turns into \eqref{current:r}.
\end{thebibliography}
\end{document}